\documentclass[12pt, preprint]{aastex}
\usepackage{amsmath}
\usepackage{graphicx}
\usepackage{natbib}

\usepackage[colorlinks=true,citecolor=blue,breaklinks=true,linktocpage=true]{hyperref}
\bibpunct{(}{)}{;}{a}{}{,} 
\usepackage{xspace}
\usepackage{hyperref}

\usepackage{multirow}

\def\vec#1{\ensuremath{\mathbf{#1}}}

\shorttitle{Kink and sausage modes in nonuniform slabs}
\shortauthors{Chen et al.}

\begin{document}


\title{KINK AND SAUSAGE MODES IN NONUNIFORM MAGNETIC SLABS WITH CONTINUOUS TRANSVERSE DENSITY DISTRIBUTIONS}

\author{Hui Yu\altaffilmark{1}}
\author{Bo Li\altaffilmark{1}}\email{bbl@sdu.edu.cn}
\author{Shao-Xia Chen\altaffilmark{1}}
\and
\author{Ming-Zhe Guo\altaffilmark{1}}
\altaffiltext{1}{Shandong Provincial Key Laboratory of Optical Astronomy and Solar-Terrestrial Environment, Institute of Space Sciences, Shandong University,Weihai, 264209, China}

\begin{abstract}
We examine the influence of a continuous density structuring transverse to coronal slabs on the dispersive properties 
    of fundamental standing kink and sausage modes supported therein.
We derive generic dispersion relations (DRs) governing linear fast waves in pressureless straight slabs with general transverse density distributions,
   and focus on the cases where the density inhomogeneity takes place in a layer of arbitrary width 
   and in arbitrary form.
The physical relevance of the solutions to the DRs is demonstrated by the corresponding time-dependent computations.
For all profiles examined, the lowest-order kink modes are trapped regardless
   of longitudinal wavenumber $k$.
A continuous density distribution introduces a difference to their periods of $\lesssim 13\%$
   when $k$ is the observed range, relative to
   the case where the density profile takes a step-function form.
Sausage modes and other branches of kink modes are leaky at small $k$,
   and their periods and damping times are heavily influenced by how the transverse density profile is prescribed,
   the lengthscale in particular.
These modes have sufficiently high quality to be observable only for
   physical parameters representative of flare loops.
We conclude that while the simpler DR pertinent to a step-function profile can be used for the lowest-order kink modes,
   the detailed information on the transverse density structuring needs to be incorporated into studies
   of sausage modes and higher-order kink modes.
\end{abstract}
\keywords{magnetohydrodynamics (MHD) --- Sun: flares --- Sun: corona --- Sun: magnetic fields --- waves}

\section{INTRODUCTION}
\label{sec_intro}
Considerable progress has been made in recent years in the field of
    solar magneto-seismology
    (see \citeauthor{2007SoPh..246....1B}~\citeyear{2007SoPh..246....1B},
    \citeauthor{2009SSRv..149....1N}~\citeyear{2009SSRv..149....1N},
    \citeauthor{2011SSRv..158..167E}~\citeyear{2011SSRv..158..167E} for three recent topical issues).
This is made possible thanks to the abundant measurements of low-frequency waves
    and oscillations in a rich variety of atmospheric structures
    on the Sun 
    (for recent reviews, see, e.g., 
    \citeauthor{2005LRSP....2....3N}~\citeyear{2005LRSP....2....3N},
    \citeauthor{2007SoPh..246....3B}~\citeyear{2007SoPh..246....3B},
    \citeauthor{2012RSPTA.370.3193D}~\citeyear{2012RSPTA.370.3193D}).
Equally important is a refined theoretical understanding of the collective waves
    in a structured magnetic environment, built on the original ideas put forward
    in the 1970s 
    {\bf and 80s
    (notably \citeauthor{1970PASJ...22..341U}~\citeyear{1970PASJ...22..341U},
    \citeauthor{1970A&A.....9..159R}~\citeyear{1970A&A.....9..159R},
    \citeauthor{1975IGAFS..37....3Z}~\citeyear{1975IGAFS..37....3Z},
    \citeauthor{1984ApJ...279..857R}~\citeyear{1984ApJ...279..857R}).}
A combination of theories and observations then enables the inference of solar atmospheric parameters
    that prove difficult to measure directly.
Take the two most-studied transverse waves in the corona, 
    the fast kink and sausage ones, for instance.
The periods of standing kink modes can offer the key information
    on the magnetic field strength 
    {\bf in coronal loops
    \citep[e.g.,][]{1984ApJ...279..857R,2001A&A...372L..53N,2008A&A...489L..49E, 2008A&A...482L...9O, 2012A&A...537A..49W}.
    }
Their damping times can help infer 
    {\bf the transverse density structuring
    \citep[e.g.,][]{2002ApJ...577..475R,2007A&A...463..333A,2008A&A...484..851G}}, 
    given that this damping
    is mostly attributable to resonant absorption
    (see \citeauthor{2011SSRv..158..289G}~\citeyear{2011SSRv..158..289G} for a review).
On the other hand, sausage modes were suggested to be responsible for causing
    a substantial fraction of quasi-periodic-pulsations (QPPs) in 
    the lightcurves of solar flares 
    \citep[e.g.,][]{2009SSRv..149..119N}.
Their periods and damping times can be employed to yield
    the magnetic field strength in the key region where flare energy is released,
    as well as the transverse density inhomogeneity of flare loops
    \citep[e.g.,][]{2003A&A...412L...7N,2012ApJ...761..134N, 2015ApJ...812...22C}.
    
Fast kink and sausage modes supported by coronal cylinders are rather well understood.
If the physical parameters are transversally structured in a piece-wise constant (step-function)
    fashion,
    the lowest-order kink modes are trapped regardless of the longitudinal wavenumber $k$,
    whereas sausage modes as well as other branches of kink modes
    are trapped only when $k$ exceeds some cut-off value
    \citep[e.g.,][]{1983SoPh...88..179E}.
In the leaky regime, the wave energy is not well confined to coronal cylinders but
    is transmitted into their
    surroundings~\citep[e.g.,][]{1982SoPh...75....3S,1986SoPh..103..277C}.
In addition, the sausage mode period (damping time) is known to increase (decrease)
    with decreasing $k$, and becomes $k$-independent when $k$ is sufficiently small
    \citep{2007AstL...33..706K,2014ApJ...781...92V}. 
If the cylinders are continuously structured in the transverse direction,
    the first branch of kink modes becomes resonantly coupled to torsional
    Alfv\'en waves and experiences temporal damping as well
    (\citeauthor{2002ApJ...577..475R}~\citeyear{2002ApJ...577..475R},
     \citeauthor{2002A&A...394L..39G}~\citeyear{2002A&A...394L..39G},
    and also \citeauthor{1988JGR....93.5423H}~\citeyear{1988JGR....93.5423H}).
Their damping times show some considerable dependence on the lengthscale, or equivalently 
    the steepness, of the transverse density distribution
    \citep[e.g.,][]{2014ApJ...781..111S}.
On the other hand, while the $k$-dependence of the sausage mode periods $P$ and damping times $\tau$
    is reminiscent of the step-function case, 
    the values of $P$~\citep{2012ApJ...761..134N,2014A&A...568A..31L,2015SoPh..tmp..118C}
    and $\tau$~\citep{2015ApJ...812...22C}  
    may be sensitive to the transverse density structuring.
Indeed, this profile dependence has inspired us \citep[][hereafter paper I]{2015ApJ...812...22C}
    to construct a scheme for inferring flare loop parameters, the transverse density lengthscale 
    in particular, with QPP measurements.

{\bf
Initiated in the comprehensive studies by \citet{1981SoPh...69...39R} and \citet{1982SoPh...76..239E},
   a considerable number of investigations into collective waves in coronal slabs
   are also available.
On the one hand, waves in slabs are easier to handle mathematically,
   and their examination can provide a useful guide to
   what one may encounter in examining waves in cylinders.
On the other hand, while waves in slabs are usually expected to be less applicable
   to observations in the solar atmosphere,
   in some situations} 
   a slab geometry {\bf was considered to be more} relevant.
For instance, 
    {\bf fast sausage waves in slabs}
    were employed to account for
    the Sunward moving tadpoles in post-flare super-arcades
    measured with the Transition Region and Coronal Explorer (TRACE)
    \citep{2005A&A...430L..65V}. 
Likewise, the large-scale propagating disturbances in streamer stalks, as seen in
    images obtained with 
    the Large Angle and Spectrometric Coronagraph (LASCO) 
    onboard the Solar and Heliospheric Observatory satellite (SOHO),
    were interpreted in terms of
    fast kink waves supported by magnetic slabs
    \citep{2010ApJ...714..644C,2011ApJ...728..147C}. 
Interestingly, 
    {\bf the theoretical results
    of fast collective waves supported by slabs
    are also applicable even in the presence of current sheets
    \citep{1986GeoRL..13..373E,1997A&A...327..377S,2011SoPh..272..119F,2014A&A...567A..24H}}.
As a matter of fact, fast sausage waves supported by current sheets were suggested 
    to be responsible for some fine structures in broadband type IV radio bursts
    \citep[e.g.,][]{2013A&A...550A...1K}.

There seems to be an apparent lack of a systematic study on
    how {\bf continuous} transverse structuring influences the dispersive properties of fast waves
    in coronal slabs.  
Besides the step-function case, only a limited number of analytical dispersion relations (DRs) are available
    for continuous density profiles of some specific form
    \citep{1988A&A...192..343E,1995SoPh..159..399N,2015ApJ...801...23L}. 
On the other hand, numerical studies from an initial-value-problem perspective 
    are primarily interested in the time signatures
    of fast waves in slabs with some prescribed continuous density profiles
    \citep{1993SoPh..143...89M,2004MNRAS.349..705N,2014A&A...567A..24H}.
In the cylindrical case, however, analytical DRs are now available for     
    transverse density distributions that are essentially arbitrary,
    and systematic investigations into the associated effects are available
    for both kink \citep{2013ApJ...777..158S,2014ApJ...781..111S}
    and sausage modes (paper I).
One naturally asks: is a similar practice possible in the slab geometry?
The present study aims to present such a practice.
To this end, we will derive an analytical DR
    governing linear fast waves
    hosted by magnetized slabs with a rather general transverse density distribution.
The only requirement here is that the density is uniform beyond some distance from a slab.
However, the density profile within this distance is allowed to be in
    arbitrary form and of arbitrary steepness.
Mathematically, this derivation largely follows paper I, and capitalizes on the fact that
    when fast waves are restricted to be in the plane containing the slab axis and
    the direction of density inhomogeneity, neither kink nor sausage waves
    resonantly couple to shear Alfv\'en waves.
Regular series expansions about some point located in the inhomogeneous part of
    the density distribution can then be used to describe fast wave perturbations.

This manuscript is organized as follows.
Section~\ref{sec_model} presents the derivation of the DR 
    together with our method of solution.
A parameter study is then presented in Sect.~\ref{sec_NumRes}
    to examine how the periods and damping times of fast waves depend on the slab parameters,
    the steepness of the transverse density profile in particular,
    for a number of different profile prescriptions.
Finally, Sect.~\ref{sec_conc} closes this manuscript with our summary
    and some concluding remarks.

\section{MATHEMATICAL FORMULATION}
\label{sec_model}
\subsection{Description for Equilibrium Slabs}
\label{sec_sub_equilibrium}
We model coronal structures as straight, density-enhanced slabs of
   half-width $R$ and aligned with a uniform magnetic field $\vec{B} = {B}\hat{z}$.
The equilibrium density $\rho$ is assumed to be inhomogeneous only in the $x$-direction,
   and $\rho(x)$ is an even function.
Apart from this, the only requirement is that $\rho$ is uniform beyond some distance 
   from the slab, making our analysis applicable to a rich variety of
   density profiles.
In the majority of our study, we examine the profiles that can be decomposed into a uniform
   {\bf core} 
   with density $\rho_{\rm i}$, a uniform external medium with density $\rho_{\rm e}$,
   and a transition layer connecting the two ($\rho_{\rm i} > \rho_{\rm e}$).
In other words,   
\begin{eqnarray}
\label{eq_rho_profile}
 {\rho}(x)=\left\{
   \begin{array}{ll}
   \rho_{\rm i},    & 0\le x \leq x_{\rm i} = R-l/2, \\
   \rho_{\rm tr}(x),& x_{\rm i} \le x \le x_{\rm e} = R+l/2,\\
   \rho_{\rm e},    & x \ge x_{\rm e} ,
   \end{array}
   \right.
\end{eqnarray}
   where the thickness ($l$) of this transition layer is bounded by
   $0$ and $2 R$, corresponding to the steepest and least steep cases, respectively.
In Appendix~\ref{sec_App_DR}, we will show that our analysis can be readily extended to profiles without 
   a uniform {\bf core}.

\begin{figure}
\centering
\includegraphics[width=0.6\columnwidth]{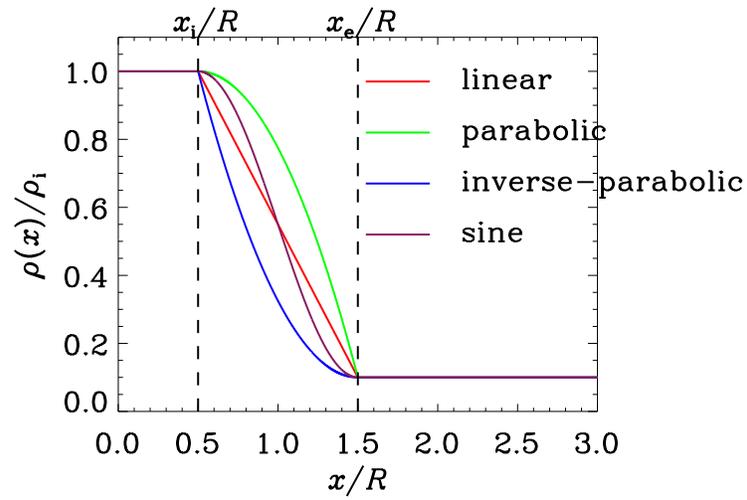}
 \caption{
 Transverse equilibrium density profiles as a function of $x$.
 The profiles differ only in how they are described in a transition layer
     connecting the internal ($\rho_{\rm i}$)
     and external ($\rho_{\rm e}$) values.
 For illustration purposes, $\rho_{\rm i}/\rho_{\rm e}$ is chosen to be $10$,
     and the width of this transition layer ($l$)
     is chosen to equal the slab half-width ($R$).
}
 \label{fig_illus_profile}
\end{figure}

Similar to Paper I, we examine the following profiles,
\begin{eqnarray}
\label{eq_rho_tr}
   \rho_{\rm tr}(x)=\left\{
   \begin{array}{ll}
   \rho_{\rm i}-\displaystyle\frac{\rho_{\rm i}-\rho_{\rm e}}{l}\left(x-x_{\rm i}\right),   & {\rm linear},
   \\[0.3cm]
   \rho_{\rm i}-\displaystyle\frac{\rho_{\rm i}-\rho_{\rm e}}{l^2}\left(x-x_{\rm i}\right)^2,& {\rm parabolic},\\[0.3cm]
   \rho_{\rm e}-\displaystyle\frac{\rho_{\rm e}-\rho_{\rm i}}{l^2}\left(x-x_{\rm e}\right)^2,    & {\rm inverse-parabolic}, \\[0.3cm]
   \displaystyle\frac{\rho_{\rm i}}{2}\left[\left(1+
   \displaystyle\frac{\rho_{\rm e}}{\rho_{\rm i}}\right)-\left(1-
   \displaystyle\frac{\rho_{\rm e}}{\rho_{\rm i}}\right)\sin\displaystyle\frac{\pi(x-R)}
   {l}\right],    & {\rm sine}.
   \end{array}
   \right.
\end{eqnarray}
{\bf We note that the sine profile was first introduced by \citet{2002ApJ...577..475R} when examining the effect of resonant
    absorption in damping standing kink modes in 
    transversally nonuniform coronal cylinders.}
Our analysis of fast waves is valid for arbitrary prescriptions of $\rho_{\rm tr}(x)$,
    the specific profiles are chosen here only to allow a quantitative analysis.
Figure~\ref{fig_illus_profile} shows the
    $x$-dependence of the chosen $\rho$ profiles,
    where for illustration purposes, 
    $\rho_{\rm i}/\rho_{\rm e}$ and $l/R$ are chosen to be
    $10$ and $1$, respectively.

\subsection{Solutions for Transverse Lagrangian Displacement and Total Pressure Perturbation}
\label{sec_sub_xirptot}
Appropriate for the solar corona, we adopt the framework of cold (zero-$\beta$) MHD.
In addition, we consider only fast waves in the $x-z$ plane
   by letting $\partial/\partial y\equiv 0$.
Let $\delta \vec{v}$ and $\delta \vec{b}$ denote 
   the velocity and magnetic field perturbations, respectively.
One finds that $\delta v_y, \delta v_z$ and $\delta b_y$ vanish. 
The perturbed magnetic pressure, or equivalently total pressure in this zero-$\beta$ case,
   is $\delta p_{\rm tot} = B\delta b_z/4\pi$.
Fourier-decomposing any perturbed value $\delta f(x, z, t)$ as
\begin{eqnarray}
\label{eq_Fourier_ansatz}
  \delta f(x,z, t)={\rm Re}\left\{\tilde{f}(x)\exp\left[-i\left(\omega t-kz\right)\right]\right\}~,
\end{eqnarray}
   one finds from linearized, ideal, cold MHD equations that
\begin{eqnarray}
\label{eq_xi}
    \displaystyle\frac{{\rm d}^2\tilde{\xi}_x}{{\rm d}x^2}
   +\left(\displaystyle\frac{\omega^2}{v^2_{\rm A}}
    -k^2\right)\tilde{\xi}_x=0~.
\end{eqnarray}
Here $\tilde{\xi}_x = i\tilde{v}_x/\omega$ is the Fourier amplitude of the transverse
    Lagrangian displacement,
    and $v_{\rm A}(x)=B/\sqrt{4\pi\rho(x)}$ is the Alfv\'en speed.
The Fourier amplitude of the perturbed total pressure is
\begin{eqnarray}
\label{eq_Fourie_ptot}
    \tilde{p}_{\rm tot} =
      -\displaystyle\frac{{B}^2}{4\pi}\displaystyle\frac{{\rm d}\tilde{\xi}_x}{{\rm d}x}~ .
\end{eqnarray}
{\bf We note that Eqs.~(\ref{eq_xi}) and (\ref{eq_Fourie_ptot}) can be derived by 
   letting the sound speeds vanish in 
   the finite-$\beta$ expressions given by 
  \citeauthor{1981SoPh...69...27R}
  (\citeyear{1981SoPh...69...27R}, Eqs.~16 and 18 therein).}

Evidently, the equations governing fast waves (see Eq.~\ref{eq_xi}) 
    do not contain any singularity.
This is different
    from the cylindrical case where a treatment of singularity is necessary to
    address the resonant coupling of kink waves to torsional Alfv\'en waves
    \citep[e.g.,][and references therein]{2013ApJ...777..158S}.
Mathematically, the solutions to Eq.~(\ref{eq_xi}) in the transition layer can be expressed as linear combinations of
    two linearly independent solutions, $\xi_{\rm tr, 1}$ and $\xi_{\rm tr, 2}$,
    that are regular series expansions about $\zeta \equiv x-R =0$.
In other words, 
\begin{eqnarray}
\label{eq_xi1xi2_expansion}
  \tilde{\xi}_{\rm tr, 1}(\zeta) = \sum_{n=0}^\infty a_n \zeta^n~, \hspace{0.2cm}
  \tilde{\xi}_{\rm tr, 2}(\zeta) = \sum_{n=0}^\infty b_n \zeta^n~.
\end{eqnarray}
{\bf Without loss of generality, one may choose 
\begin{eqnarray}
\label{eq_a01_b01} 
    a_0 = R, \hspace{0.2cm} a_1 = 0, \hspace{0.2cm} 
    b_0 = 0, \hspace{0.2cm} b_1 = 1. 
\end{eqnarray} }
To determine the coefficients $[a_n, b_n]$ for $n\ge 2$, 
    we expand $\rho_{\rm tr}(x)$ about $\zeta =0$ as well, resulting in
\begin{eqnarray}
\label{eq_rho_expansion}
    \rho_{\rm tr}(\zeta) = \rho_0+\sum^\infty_{n=1}\rho_n \zeta^n,
\end{eqnarray}
    where 
\begin{eqnarray}
\label{eq_rho_coef}
  \rho_0 = \rho|_{\zeta=0}, \mbox{ and }
  \rho_n = \frac{1}{n!} \left.\frac{{\rm d}^n\rho(\zeta)}{{\rm d}\zeta^n}\right|_{\zeta=0}
     \hspace{0.2cm} (n\ge 1).
\end{eqnarray}
Now substituting Eq.~(\ref{eq_xi1xi2_expansion}) into Eq.~(\ref{eq_xi})
    with the change of independent variable from
    $x$ to $\zeta$, one arrives at
\begin{eqnarray}
\label{eq_a_n}
 && \chi_n = -\displaystyle\frac{1}{n(n-1)}\left[\frac{4\pi\omega^2}{B^2}
  \displaystyle\sum\limits_{l=0}^{n-2}\rho_{n-2-l}\chi_l-k^2\chi_{n-2}\right]~~~~~
  ~~(n\ge2) ,
\end{eqnarray}
    where $\chi$ represents either $a$ or $b$.
    
The Fourier amplitude of the transverse Lagrangian displacement reads
\begin{eqnarray}
\label{eq_xi_solution_entire}
   \tilde{\xi}_x(x)=\left\{
   \begin{array}{ll}
   \left\{
   \begin{array}{ll}
   A_{\rm i}\sin(\mu_{\rm i} x)~,& {\rm sausage}\\[0.1cm]
   A_{\rm i}\cos(\mu_{\rm i} x)~,& {\rm kink}
   \end{array}
   \right.,                         & 0 \le x \le x_{\rm i}, \\ [0.1cm]
   A_1\tilde{\xi}_{\rm tr,1}(\zeta)
  +A_2\tilde{\xi}_{\rm tr,2}(\zeta), & x_{\rm i} < x < x_{\rm e},\\[0.1cm]
   A_{\rm e}\exp(i\mu_{\rm e} x),   & x \ge x_{\rm e},
   \end{array}
   \right.
\end{eqnarray}
    where $A_{\rm i}, A_{\rm e}, A_1$ and $A_2$ are arbitrary constants.
In addition, 
\begin{eqnarray}
\label{eq_def_muie} 
   \mu_{\rm i, e} = \sqrt{\frac{\omega^2}{v_{\rm Ai, e}^2} - k^2} 
   \hspace{0.4cm} \left(-\frac{\pi}{2} < \arg\mu_{\rm i,e} \le \frac{\pi}{2}\right),
\end{eqnarray}
    with $v_{\rm Ai, e}^2 = {B}^2/(4\pi\rho_{\rm i, e})$. 
With the aid of Eq.~(\ref{eq_Fourie_ptot}), the Fourier amplitude of the total
    pressure perturbation evaluates to
\begin{eqnarray}
   \tilde{p}_{\rm tot}(x)=-\displaystyle\frac{{B}^2}{4\pi}\times\left\{
   \begin{array}{ll}
   \left\{
   \begin{array}{ll}
	A_{\rm i}\mu_{\rm i}\cos(\mu_{\rm i} x)~,& {\rm sausage}\\[0.1cm]
       -A_{\rm i}\mu_{\rm i}\sin(\mu_{\rm i} x)~,& {\rm kink}
   \end{array}
   \right.,    & 0 \le x \le x_{\rm i}, \\ [0.1cm]
   A_1\tilde{\xi}_{\rm tr,1}^\prime(\zeta)
  +A_2\tilde{\xi}_{\rm tr,2}^\prime(\zeta),	 & x_{\rm i} < x < x_{\rm e},\\[0.1cm]
   iA_{\rm e}\mu_{\rm e}\exp(i\mu_{\rm e} x),    & x \geq x_{\rm e},
   \end{array}
   \right.
\end{eqnarray}
    where the prime $'={\rm d}/{\rm d}\zeta$.
    
A few words are necessary to address the restriction on $\arg\mu_{\rm i, e}$.
As will be obvious in the derived DR, allowing $\mu_{\rm i}$ to take
    the negative square root in Eq.~(\ref{eq_def_muie}) will not introduce 
    additional independent solutions.
On the other hand, the restriction on $\arg{\mu_{\rm e}}$ excludes the unphysical solutions
    that correspond to purely growing perturbations (with $\mu_{\rm e}$ lying on the negative imaginary axis)
    in the external
    medium~\citep[see][for a discussion on these unphysical solutions]{2005A&A...441..371T}.
This restriction also allows a unified examination of both trapped and leaky waves. 
Indeed, the trapped regime arises when $\arg\mu_{\rm e} = \pi/2$, in which case one finds that
    $\exp(i\mu_{\rm e} x) = \exp(-|\mu_{\rm e}| x)$.    
For a similar discussion in the cylindrical case, see \citet{1986SoPh..103..277C}.

\subsection{Dispersion Relations of Fast Waves}
\label{sec_sub_DR}
The dispersion relations (DR) governing linear fast waves follow from
     the requirement that both $\tilde{\xi}_x$ and $\tilde{p}_{\rm tot}$
     be continuous at the interfaces
     $x=x_{\rm i}$ and $x=x_{\rm e}$.
This leads to
\begin{eqnarray*}
 A_1\tilde{\xi}_{\rm tr,1}(\zeta_{\rm i})
    +A_2\tilde{\xi}_{\rm tr,2}(\zeta_{\rm i})&=& \left\{
   \begin{array}{ll}
   A_{\rm i}\sin(\mu_{\rm i} x_{\rm i})~,& {\rm sausage}\\[0.1cm]
   A_{\rm i}\cos(\mu_{\rm i} x_{\rm i})~,& {\rm kink}
   \end{array}
   \right.,    \\
  A_1\tilde{\xi}_{\rm tr,1}(\zeta_{\rm e})+A_2\tilde{\xi}_{\rm tr,2}(\zeta_{\rm e})&=& A_{\rm e}\exp(i\mu_{\rm e} x_{\rm e}) \\ [0.1cm]
  A_1\tilde{\xi}_{\rm tr,1}^\prime(\zeta_{\rm i})+A_2\tilde{\xi}_{\rm tr,2}^\prime(\zeta_{\rm i})&=& \left\{
   \begin{array}{ll}
    A_{\rm i}\mu_{\rm i}\cos(\mu_{\rm i} x_{\rm i})~,&  {\rm sausage}\\[0.1cm]
   -A_{\rm i}\mu_{\rm i}\sin(\mu_{\rm i} x_{\rm i})~,& {\rm kink}
   \end{array}
   \right.,
   \\[0.1cm]
  A_1\tilde{\xi}_{\rm tr,1}^\prime(\zeta_{\rm e})+A_2\tilde{\xi}_{\rm tr,2}^\prime(\zeta_{\rm e})&=& iA_{\rm e}\mu_{\rm e}\exp(i\mu_{\rm e} x_{\rm e})
\end{eqnarray*}
    where $\zeta_{\rm i}=-l/2$ and $\zeta_{\rm e}=l/2$.
Eliminating $A_{\rm i}$ ($A_{\rm e}$), one finds that
\begin{equation}
\label{eq_prep_DR}
\begin{array}{rcl}
&&  \Lambda_1 A_1 + \Lambda_2 A_2 = 0,  \\
&&  \Lambda_3 A_1 + \Lambda_4 A_2 = 0,
\end{array}
\end{equation}
   where the coefficients read
\begin{eqnarray}
\label{eq_Ys}
\begin{array}{rcl}
   \Lambda_1=X_{\rm i}\tilde{\xi}_{\rm tr,1}(\zeta_{\rm i})-\tilde{\xi}_{\rm tr,1}^\prime(\zeta_{\rm i}) , 
      & 
   \Lambda_2=X_{\rm i}\tilde{\xi}_{\rm tr,2}(\zeta_{\rm i})-\tilde{\xi}_{\rm tr,2}^\prime(\zeta_{\rm i}) , \\
   \Lambda_3=X_{\rm e}\tilde{\xi}_{\rm tr,1}(\zeta_{\rm e})-\tilde{\xi}_{\rm tr,1}^\prime(\zeta_{\rm e}) , 
      & 
   \Lambda_4=X_{\rm e}\tilde{\xi}_{\rm tr,2}(\zeta_{\rm e})-\tilde{\xi}_{\rm tr,2}^\prime(\zeta_{\rm e}) ,
\end{array}
\end{eqnarray}
with
\begin{eqnarray}
\label{eq_XX}
\begin{array}{rcl}
   X_{\rm e} = i\mu_{\rm e}, \hspace{0.2cm}
   X_{\rm i} = \left\{
   \begin{array}{ll}
   \mu_{\rm i}\cot(\mu_{\rm i} x_{\rm i})~,&  {\rm sausage}\\[0.1cm]
   -\mu_{\rm i}\tan(\mu_{\rm i} x_{\rm i})~,& {\rm kink}
   \end{array}
   \right.
\end{array}
\end{eqnarray}
Evidently, for Eq.~(\ref{eq_prep_DR}) to allow non-trivial solutions
of $[A_1, A_2]$, one needs to require that
\begin{eqnarray}
\label{eq_DR} \Lambda_1 \Lambda_4 - \Lambda_2 \Lambda_3 = 0 ,
\end{eqnarray}
   which is the DR we are looking for.

In the limit $l/R\rightarrow 0$, Eq.~(\ref{eq_DR}) should 
    recover the well-known results for the step-function profile.
This can be readily shown by retaining only terms to the 0-th order in $l/R$ and by noting that
    $x_{\rm i}\approx x_{\rm e} \approx R$.
The coefficients $\Lambda_{n}~(n=1, \cdots, 4)$ simplify to
\begin{eqnarray*}
&&  \Lambda_1=X_{\rm i}a_0-a_1  ,    \\
&&  \Lambda_2=X_{\rm i}b_0 -b_1 ,  \\
&&  \Lambda_3=X_{\rm e}a_0-a_1  ,  \\
&&  \Lambda_4=X_{\rm e}b_0-b_1  .
\end{eqnarray*}
Substituting these expressions into Eq.~(\ref{eq_DR}), one finds
that
\begin{eqnarray*}
(X_{\rm i}-X_{\rm e})(a_1 b_0-a_0 b_1)=0 .
\end{eqnarray*}
Now that $a_1 b_0 - a_0 b_1$ cannot be zero since
    $\tilde{\xi}_{\rm tr, 1}$ and $\tilde{\xi}_{\rm tr, 2}$ are linearly independent, 
    one finds that $X_{\rm i} = X_{\rm e}$. 
To be specific (see Eq.~\ref{eq_XX}),
\begin{eqnarray}
\label{eq_DR_tophat}
  \left\{
  \begin{array}{ll}
      i\mu_{\rm e} =  \mu_{\rm i}\cot(\mu_{\rm i} R)~,&  {\rm sausage}\\[0.1cm]
      i\mu_{\rm e} = -\mu_{\rm i}\tan(\mu_{\rm i} R)~,& {\rm kink}
  \end{array}
  \right.
\end{eqnarray}
   which is the DR for step-function density profiles \citep[e.g.,][]{2005A&A...441..371T,2013ApJ...767..169L}.

Before proceeding, we note that the DR is valid for both propagating and standing waves.
Throughout this study, however, we focus on standing modes by assuming that
    the longitudinal wavenumber $k$ is real, whereas
    the angular frequency $\omega$ can be complex-valued.
In addition, wherever applicable, we examine only fundamental modes, namely, the 
    modes with $k=\pi/L$ where $L$ is the slab length.
Once a choice for $\rho_{\rm tr}$ is made,
    $\omega$ in units of $v_{\rm Ai}/R$ depends only on the dimensionless parameters $[\rho_{\rm i}/\rho_{\rm e}, l/R, kR]$.
A numerical approach is then necessary to solve the DR (Eq.~\ref{eq_DR}).
For this purpose,
{\bf 
    we start with evaluating the coefficients $\rho_n$ with Eq.~(\ref{eq_rho_coef}),
    and then evaluate $a_n$ and $b_n$ with Eq.~(\ref{eq_a_n}).
The coefficients $\Lambda_n$ can be readily obtained with Eqs.~(\ref{eq_Ys})
    and (\ref{eq_XX}),
    thereby allowing us to solve Eq.~(\ref{eq_DR}).
When evaluating $\Lambda_n$, we truncate the infinite series expansion in Eq.~(\ref{eq_xi1xi2_expansion}) 
    by retaining only the terms with $n$ up to $N=101$.
}
A convergence test has been made for a substantial fraction of the numerical results
    to make sure that using an even larger $N$ does not introduce any appreciable difference.

{\bf 
Given that a numerical approach is needed after all, one may ask why not treat the problem numerically from
    the outset.
Let us address this issue by comparing our approach
    with two representative fully numerical methods
    for examining the dispersive properties of fast modes.
First, one may solve Eq.~(\ref{eq_xi}) as an eigenvalue problem with a chosen $v_{\rm A}(x)$ profile 
    \citep[e.g.,][]{2007SoPh..246..165P,2012A&A...537A..46J,2014A&A...568A..31L}.
However, this approach usually needs to specify the outer boundary condition 
    that the Lagrangian displacement $\tilde{\xi}$ vanishes,
    and therefore can be used only to examine trapped modes since 
    $\tilde{\xi}$ diverges rather than vanishing for leaky modes 
    at large distances.
Second, one may obtain a time-dependent equation governing, say, the transverse velocity perturbation
    and find the periods and damping times of fast modes
    by analyzing the temporal evolution of the perturbation signals
    (see Appendix~\ref{sec_App_IVP} for details).
Compared with this approach, solving the analytical DR is substantially less 
    computationally expensive, 
    thereby allowing an exhaustive parameter study to be readily conducted.
On top of that, the periods and damping times for heavily damped modes
    can be easily evaluated, whereas
    the perturbation signals may decay too rapidly to permit
    their proper determination.
}    

For future reference, we note that Eq.~(\ref{eq_DR_tophat}) allows one to derive explicit expressions
    for $\omega$ at $k=0$ and for the critical wavenumber $k_{\rm c}$
    that separates the leaky from trapped regimes
    \citep[][Eqs.~8 to 13]{2005A&A...441..371T}. 
With our notations, the results for sausage modes can be expressed as   
\begin{eqnarray}
\label{eq_step_saus_k0}
 \begin{array}{ll}
 k_{\rm c} = \displaystyle\frac{1}{R}\displaystyle\frac{(n+1/2)\pi}{\sqrt{v_{\rm Ae}^2/v_{\rm Ai}^2-1}}, \\[0.4cm]
 \omega_{\rm R} =  \displaystyle\frac{v_{\rm Ai}}{R}\left(n+\frac{1}{2}\right)\pi , \\[0.4cm]
 \omega_{\rm I} = -\displaystyle\frac{v_{\rm Ai}}{2R}\ln\displaystyle\frac{1+v_{\rm Ai}/v_{\rm Ae}}{1-v_{\rm Ai}/v_{\rm Ae}} , 
 \end{array}
\end{eqnarray}
   where $n=0, 1, \cdots$.
Similarly, the results for kink modes read
\begin{eqnarray}
\label{eq_step_kink_k0}
 \begin{array}{ll}
 k_{\rm c} = \displaystyle\frac{1}{R}\displaystyle\frac{n\pi}{\sqrt{v_{\rm Ae}^2/v_{\rm Ai}^2-1}}, \\[0.4cm]
 \omega_{\rm R} =  \displaystyle\frac{v_{\rm Ai}}{R}n\pi , \\[0.4cm]
 \omega_{\rm I} = -\displaystyle\frac{v_{\rm Ai}}{2R}\ln\displaystyle\frac{1+v_{\rm Ai}/v_{\rm Ae}}{1-v_{\rm Ai}/v_{\rm Ae}} , 
 \end{array}
\end{eqnarray}
   where $n=1, 2, \cdots$.
   
\section{NUMERICAL RESULTS}
\label{sec_NumRes}
\subsection{Overview of Dispersion Diagrams}
\label{sec_sub_physrel}

Let us start with an overview of the dispersion diagrams 
    representing the solutions to the DR.
Figure~\ref{fig_disp_diag} presents the
    dependence on longitudinal wavenumber $k$ of
    the real ($\omega_{\rm R}$, the upper row)
    and imaginary ($\omega_{\rm I}$, lower)
    parts of angular frequency for 
    kink (the left column) and sausage (right) modes.
Note that $-\omega_{\rm I}$ is plotted instead of $\omega_{\rm I}$ since $\omega_{\rm I} \le 0$.    
For illustration purposes, a combination of $[\rho_{\rm i}/\rho_{\rm e}, l/R]=[10, 1]$
    is chosen.
Four different choices of density profiles are presented in different colors
    as labeled, and the results pertinent to the step-function case
    are presented by the black solid curves for comparison.
The dash-dotted lines in Figs.~\ref{fig_disp_diag}a and \ref{fig_disp_diag}c 
    represent $\omega_{\rm R} = k v_{\rm Ae}$
    and separate the trapped (to their right) from leaky modes (left). 
There are an infinite number of solutions given the transcendental nature of Eq.~(\ref{eq_DR}).
Therefore for kink modes we choose to examine only the first two branches
    (labeled I and II),
    whereas for sausage modes we examine only the first one (labeled I).

Consider kink modes first.
The branches labeled I in Fig.~\ref{fig_disp_diag}a always lie below the dash-dotted line, 
    and the associated $\omega_{\rm I}$ is zero  (not shown).
Hence these solutions pertain to trapped modes, regardless of longitudinal wavenumber.
Comparing the black solid line with those in various colors,
    one finds that for the parameters examined here,
    the difference introduced by 
    replacing the step-function profile with the examined continuous profiles 
    seems marginal.
As for the branches labeled II,
    one sees from Fig.~\ref{fig_disp_diag}a that common to all profiles,
    $\omega_{\rm R}$ monotonically decreases     
    with decreasing $k$, and its $k$-dependence
    in the leaky regime 
    is substantially weaker
    than in the trapped one.
Likewise, 
    Fig.~\ref{fig_disp_diag}b indicates that regardless of specific profiles, 
    $|\omega_{\rm I}|$ monotonically increases with decreasing $k$ once entering the leaky regime.
However, the specific values of $\omega_{\rm R}$ and $\omega_{\rm I}$ for branches II
    show some considerable profile dependence.
In the leaky regime, Fig.~\ref{fig_disp_diag}a indicates that $\omega_{\rm R}$ may be larger 
    or less than in the step-function case, which occurs in conjugation
    with the changes in the critical longitudinal wavenumbers ($k_{\rm c}$) corresponding to
    the intersections between the solid curves and the dash-dotted one.
For instance, one finds that for the parabolic (inverse-parabolic) profile,  
    $\omega_{\rm R}R/v_{\rm Ai}$ reads $2.6$ ($3.51$) when $kR \rightarrow 0$,
    while $k_{\rm c} R$ attains $0.86$ ($1.19$).
These values are substantially different from those in the step-function case, where
    $\omega_{\rm R} R/v_{\rm Ai} $ attains $\pi$ at $kR \rightarrow 0$,
    and $k_{\rm c}R$ is found to be $1.05$ (see Eq.~\ref{eq_step_kink_k0}).
Examining Fig.~\ref{fig_disp_diag}b, one sees that the profile dependence of
    $\omega_{\rm I}$ is even more prominent.
Still take the values at $kR \rightarrow 0$ for example.
One finds that $\omega_{\rm I}R/v_{\rm Ai}$ attains $-0.33$ in the step-function case,
    but reads $-0.61$ when the inverse-parabolic profile is chosen.

Now consider sausage modes given in the right column.
Interestingly, 
    the overall dependence of $\omega$ on $k$
    is similar to the one for kink modes.
In particular, the $k$-dependence for continuous profiles is 
    reminiscent of the step-function case.
However, choosing different density profiles has a considerable impact on
    the specific values of both $\omega_{\rm R}$
    and $\omega_{\rm I}$.
Examine $k\rightarrow 0$ for instance.
While in the step-function case $[\omega_{\rm R}, -\omega_{\rm I}]R/v_{\rm Ai}$ is 
    $[\pi/2, 0.33]$ (see Eq.~\ref{eq_step_saus_k0}), 
    it attains $[1.33, 0.3]$ and $[1.83, 0.45]$
    for the parabolic and inverse-parabolic profiles, respectively.

As discussed in \citet{2007SoPh..246..231T} pertinent to the cylindrical case
    with a step-function density profile, 
    not all solutions in an eigen-mode analysis
    have physical relevance.
Similar to that study (also see \citeauthor{2005A&A...441..371T}~\citeyear{2005A&A...441..371T}),
    we approach this issue by solving the relevant time-dependent equations
    and then asking whether the solutions given in Fig.~\ref{fig_disp_diag}
    are present in the temporal evolution of the perturbations.
In order not to digress from the eigen-mode analysis too far,
    let us present the details in Appendix~\ref{sec_App_IVP}, 
    and simply remark here that the solutions to the dispersion relations
    as presented in Fig.~\ref{fig_disp_diag} are all physically relevant.

\subsection{Standing Kink Modes}
\label{sec_sub_kink}
This section focuses on how different choices of the $\rho_{\rm tr}$ profile impact the
    dispersive properties of standing kink modes.
Let us start with an examination of this influence on the modes labeled I in Fig.~\ref{fig_disp_diag},
    which is trapped for arbitrary longitudinal wavenumber $k$.
To proceed, we note that for typical active region (AR) loops, the ratio of half-width to length 
    ($R/L$) tends to be $\lesssim 0.1$~\citep[e.g., Fig.~1 in][]{2007ApJ...662L.119S}.
Even for relatively thick flare loops, $R/L$ tends not to exceed $0.2$
    \citep{2003A&A...412L...7N,2004ApJ...600..458A}.
We therefore ask how much the periods $P$ for various density profiles
    can deviate from the step-function counterpart ($P^{\rm step}$)
    by surveying $kR = \pi R/L$ in the range between $0$ and $0.2\pi$.
Let $\delta P$ denote
    the most significant value attained by $P/P^{\rm step}-1$
    in this range of $kR$.
Figure~\ref{fig_maxdevP_trappedkink} presents $\delta P$ as a function of $l/R$ for a number
    of density contrasts $\rho_{\rm i}/\rho_{\rm e}$ as labeled.
The results for different density profiles are presented in different panels.
One sees that the sign of $\delta P$ critically depends on the prescription
    of the density profile.
While $P$ is larger than in the step-function case for the parabolic profile,
    it is smaller for the rest of the density profiles.
In addition, $|\delta P|$ tends to increase with increasing 
    $\rho_{\rm i}/\rho_{\rm e}$ when $l/R$ is fixed. 
This tendency somehow levels off 
    when $\rho_{\rm i}/\rho_{\rm e}$ is large, as evidenced by the fact that
    the results for $\rho_{\rm i}/\rho_{\rm e}=10$
    are close to those for $\rho_{\rm i}/\rho_{\rm e}=100$.     
Nonetheless, $|\delta P|$ is consistently smaller than $13.2\%$.
As a matter of fact, when linear and sine profiles are adopted, $|\delta P|$ is
    no larger than $5.3\%$.
From this we conclude that at least for the prescriptions we adopt for the equilibrium density profile,
    one can use the simpler dispersion relation for the step-function case
    to describe the periods of the first branch of kink modes.

What will be the influence of density profiles on the kink modes
    labeled II in Fig.~\ref{fig_disp_diag}?
Figure~\ref{fig_contour_kinkPtau} quantifies this influence by presenting the distribution
    in the $[l/R, \rho_{\rm i}/\rho_{\rm e}]$ plane
    of the period $P$ (the left column)
       and damping-time-to-period ratio
    $\tau/P=\omega_{\rm R}/|2\pi\omega_{\rm I}|$ (right), both evaluated at $k=0$.
Here $\rho_{\rm i}/\rho_{\rm e}$ is taken to be in the range $[2, 200]$, encompassing
    the values for both AR loops and flare loops.
And $\tau/P$ is examined in place of $\tau$ since it is a better measure of signal quality.    
Each row represents one of the four examined density profiles,
    and the red contours in the right column represents where $\tau/P=3$, the nominal value
    required for a temporally damping signal to be measurable.
Consider the left column first.
When $l/R\rightarrow 0$, $P$ attains $2$ in all cases as expected
    for the step-function case where
    $\omega_{\rm R}=\pi$ regardless of density contrasts (see Eq.~\ref{eq_step_kink_k0}).
However, some considerable difference appears when continuous profiles are adopted.
The period $P$ may be as large as $3.2$ for the parabolic profile
    (the lower right corner in Fig.~\ref{fig_contour_kinkPtau}b).
It may also be as small as $1.76$, attained for the inverse-parabolic profile
    (the upper middle part in Fig.~\ref{fig_contour_kinkPtau}c).
The $l/R$-dependence of $P$ is also sensitive to the choice of profiles.
With the exception of the inverse-parabolic profile, 
    $P$ monotonically increases with increasing $l/R$ at a fixed density contrast.
However, when the inverse-parabolic profile is chosen,
    $P$ shows a nonmonotical dependence on $l/R$, decreasing with $l/R$ first before increasing again. 
Now examine the right column.
One sees that regardless of profiles, $\tau/P$ decreases monotonically with $l/R$ 
    when $\rho_{\rm i}/\rho_{\rm e}$ is fixed.
This is intuitively expected since more diffuse slabs should be less efficient in
    trapping wave energy.
Nevertheless, the values of $\tau/P$ are considerably different for different choices of profiles.
Compare the parabolic and inverse-parabolic profiles,
    and examine the intersections between the red curve and the vertical line representing $l/R=2$ for instance.
While for the parabolic profile this intersection corresponds to 
    $\rho_{\rm i}/\rho_{\rm e}=81$,
    a value of $\rho_{\rm i}/\rho_{\rm e}=227$ is found for the inverse-parabolic distribution.
Actually this value is beyond the range we adopt for plotting Fig.~\ref{fig_contour_kinkPtau}.
    
A question now arises: under what conditions can branch II 
    have sufficiently high quality to be observable?
As shown in Fig.~\ref{fig_disp_diag},  with decreasing $kR$, $\omega_{\rm R}$ ($|\omega_{\rm I}|$)
    decreases (increases) monotonically, meaning that $\omega_{\rm R}/|\omega_{\rm I}|$
    and consequently $\tau/P$ decrease monotonically.
This suggests that for $\tau/P$ to exceed a given value $(\tau/P)_{\rm obs}$, taken here to be $3$
    as required by observations to discern a temporally decaying signal,
    $R/L = kR/\pi$ has to exceed 
    some critical value $(R/L)_{\rm obs}$.
Figure~\ref{fig_RoLobs_kink} shows the contours
    of $(R/L)_{\rm obs}$ in the $[l/R, \rho_{\rm i}/\rho_{\rm e}]$ plane 
    for different density profiles as given in different panels.
In each panel, a contour represents the lower limit of $\rho_{\rm i}/\rho_{\rm e}$ for a given $l/R$
    for the signals associated with branch II
    to be observable when the half-width-to-length ratio $R/L$ is 
    smaller than the value given by this contour.
Put in another way, branch II is observable in the area below a contour
    only when $R/L$ exceeds at least the value represented by this contour.
The red line represents where $(\tau/P)(k=0) = 3$, meaning that branch II is always observable in 
    the hatched area bounded from below by this red curve. 
Evidently, this hatched area corresponds to high density contrasts exceeding at least $36.7$,
    which is attained when $l/R\rightarrow 0$ (see Eq.~\ref{eq_step_kink_k0}).
These high density contrasts are not unrealistic but lie in the observed range
    of flare or post-flare loops, for which $\rho_{\rm i}/\rho_{\rm e}$ 
    may reach up to $10^3$~\citep[e.g.,][]{2004ApJ...600..458A}.
The lower hatched area corresponds to density contrasts characteristic of
    AR loops, for which $2 \le \rho_{\rm i}/\rho_{\rm e} \le 10$~\citep[e.g.,][]{2004ApJ...600..458A}.
One sees that in this area, for branch II to be observable, the magnetic structures
    are required to have an $R/L$ exceeding at least $0.2$ (see the parabolic profile).
In reality, however, $R/L$ for AR loops is $\lesssim 0.1$~\citep[e.g.,][]{2007ApJ...662L.119S}.
Therefore we conclude that branch II is observationally discernible
    only for flare or post-flare loops.

\subsection{Standing Sausage Modes}
\label{sec_sub_saus}
Now move on to the sausage modes.
Figure~\ref{fig_contour_sausPtau} presents, in the same format as Fig.~\ref{fig_contour_kinkPtau},
    the periods ($P$, the left column)
    and damping-time-to-period ratios ($\tau/P$, right)
    at $k=0$ 
    for various density profiles as given in different rows. 
The red curves in the right column correspond to $\tau/P=3$.    
Consider the left column first.
One sees that with the exception of the inverse-parabolic profile, 
    $P$ is consistently larger than in the step-function case where $P=4~R/v_{\rm Ai}$ (see Eq.~\ref{eq_step_saus_k0}).
It may reach up to $6.09~R/v_{\rm Ai}$
    for the parabolic profile (the lower right corner in Fig.~\ref{fig_contour_sausPtau}b).
Furthermore, for these profiles $P$ monotonically increases with $l/R$.
In other words, the period of sausage modes increases when a slab becomes more diffuse,
    which agrees with \citet{2014A&A...567A..24H} where
    a specific continuous density profile is chosen.
However, this tendency is not universally valid.
For instance, for the inverse-parabolic profile, $P$ decreases with $l/R$ in a substantial
    fraction of the parameter space.
In this case, $P$ is consistently smaller than in the step-function case
    and reaches $3.14~R/v_{\rm Ai}$ at the upper right corner in Fig.~\ref{fig_contour_sausPtau}c.
Examining the right column, one finds that while $\tau/P$ monotonically decreases with $l/R$
    at any given density contrast for all profiles, 
    the specific values of $\tau/P$ show some considerable profile dependence.    
Compare the parabolic and inverse-parabolic profiles
    and examine the intersections between the red curves and the horizontal lines 
    representing $\rho_{\rm i}/\rho_{\rm e} = 200$.
One finds that this intersection takes place at an $l/R$ of $1.96$ for the parabolic profile,
    whereas it is located at $l/R = 1.02$ for the inverse-parabolic one.

One may also question whether the sausage modes can be observed.
Similar to Fig.~\ref{fig_RoLobs_kink}, Figure~\ref{fig_RoLobs_saus}
    presents the distribution of $(R/L)_{\rm obs}$ 
    in the $[l/R, \rho_{\rm i}/\rho_{\rm e}]$ space, where $(R/L)_{\rm obs}$
    represents the half-width-to-length ratio at which $\tau/P = 3$.
Note that the red curves represent where $(\tau/P)(k=0)=3$, meaning that     
    slabs with parameters in the hatched part bounded from below by a red curve
    always support sausage modes with sufficiently high quality, irrespective
    of their widths or lengths.
One sees that this area
    corresponds to density contrasts that are even higher than for the kink modes. 
Even when $l/R \rightarrow 0$, $\rho_{\rm i}/\rho_{\rm e}$ needs to be larger
    than $144$ (see Eq.~\ref{eq_step_saus_k0}, and note that
    $\tau/P \approx \sqrt{\rho_{\rm i}/\rho_{\rm e}}/4$ for large $\rho_{\rm i}/\rho_{\rm e}$).
However, the severe restriction on $\rho_{\rm i}/\rho_{\rm e}$ is alleviated 
    given the finite $R/L$ for flare and post-flare loops.
Consider the worst-case scenario, which takes place for the inverse-parabolic profile.
If a slab corresponds to $R/L = 0.2$, then $\rho_{\rm i}/\rho_{\rm e}$ is required to be larger than
    only $\sim 21$, which actually lies close to the lower limit of density ratios
    measured for flare loops~\citep[e.g.,][]{2004ApJ...600..458A}.
When it comes to density contrasts characteristic of AR loops, represented by the lower hatched area,
    one finds that $R/L$ has to be consistently larger than $0.1$ for the sausage modes to be observable. 
This value, however, is beyond the upper limit of the ratios of half-width to length
    for AR loops.
We therefore conclude that the sausage modes are observable only for flare or post-flare loops.

\section{CONCLUSIONS}
\label{sec_conc}

How plasma density is structured across various magnetic structures in the solar corona
    remains largely unknown.
{\bf 
The present study was intended to examine
    the influence of continuous
    transverse density structuring on fast kink and sausage modes
    collectively supported by coronal slabs.
To this end,}
    we worked in the framework of linearized, ideal, cold (zero-$\beta$) MHD and modeled
    coronal loops as straight slabs with a rather general transverse 
    density profile, the only requirement being that
    the density is uniform beyond some distance from the slab.
Analytical dispersion relations (DR) governing both fast kink and sausage waves
    were derived by solving the perturbation equations in terms of
    regular series expansions in the nonuniform part of the density distribution.
The solutions to the DRs were numerically found,
    and they were shown to be physically relevant
    in that they are present in the associated time-dependent computations.
While one class of density profiles was examined in detail where 
    a transition layer connects a uniform {\bf core} and an external uniform medium,
    we showed that a similar analysis is straightforward for density profiles without a uniform {\bf core}.

Focusing on fundamental standing modes,
    we found that their periods $P$ (and damping times $\tau$ if the modes are leaky)
    in units of $R/v_{\rm Ai}$
    depend on a combination of dimensionless parameters
    $[kR, l/R, \rho_{\rm i}/\rho_{\rm e}]$
    once a description for the transverse density profile is prescribed.
Here $R$ denotes the slab half-width, $v_{\rm Ai}$ is the Alfv\'en speed at the slab axis,
    $k$ is the longitudinal wavenumber, $l$ is the transverse density lengthscale,
    and $\rho_{\rm i}/\rho_{\rm e}$ is the density contrast between the slab and its surrounding fluid.
The lowest-order kink modes are trapped for arbitrary $k$.
For the profiles examined, their periods differ by $\lesssim 13\%$ from the case 
    where the transverse density distribution
    is a step-function one when $k$ is in the observed range.
However, sausage modes and other branches of kink modes are leaky at small $k$,
    and their periods and damping times are sensitive to 
    the choice of the transverse density profile, the density lengthscale in particular.
We also found that these fast modes have sufficiently high quality to be observable 
    only for parameters representative of flare loops.

While the transverse density distribution is allowed to be arbitrary,
    our analysis nonetheless has a number of limitations.
First, we neglected wave propagation out of the plane formed by the slab axis and the direction
    of density inhomogeneity and therefore cannot address the 
    possible coupling to shear Alfv\'en waves~\citep[e.g.,][]{1997A&A...324.1170T}. 
Second, the adopted ideal MHD approach excluded the possible role
    of dissipative mechanisms like magnetic resistivity and ion viscosity 
    in damping the fast modes.
In the cylindrical case, resistivity is known to be important for kink modes only 
    in the layer where resonant coupling takes place, and it influences only the 
    detailed structure of the eigen-functions instead of the damping time or
    period~\citep[e.g.,][]{2013ApJ...777..158S}.
On the other hand, ion viscosity or electron heat conduction is unlikely to be important for at least some
    observed sausage modes~\citep[e.g.,][]{2007AstL...33..706K}.
For wave modes in coronal slabs, ion viscosity     
    was shown to be effective only for 
    waves with frequencies exceeding several Hz \citep{1994ApJ...435..502P}.
Third, this study neglected the longitudinal variations in plasma density or magnetic field strength.
While these effects are unlikely to be important for sausage modes~\citep{2009A&A...494.1119P},
    they may need to be addressed for kink ones, especially 
    when the period ratios between the fundamental mode and its harmonics
    are used for seismological purposes~\citep[e.g.,][]{2007A&A...471..999D}

\acknowledgments
{\bf We thank the referee for the constructive comments.
This work is supported by the 973 program 2012CB825601,
} 
    the National Natural Science Foundation of China (41174154, 41274176, and 41474149),
    and by the Provincial Natural Science Foundation of Shandong via Grant JQ201212.

\bibliographystyle{apj}
\bibliography{slab}

\IfFileExists{\jobname.bbl}{} {\typeout{}
\typeout{****************************************************}
\typeout{****************************************************}
\typeout{** Please run "bibtex \jobname" to obtain} \typeout{**
the bibliography and then re-run LaTeX} \typeout{** twice to fix
the references !}
\typeout{****************************************************}
\typeout{****************************************************}
\typeout{}}

\begin{center}
{\bf APPENDIX}
\end{center}

\appendix
\section{FAST WAVES IN NON-UNIFORM SLABS: AN INITIAL-VALUE-PROBLEM APPROACH}
\label{sec_App_IVP}

This section demonstrates the physical relevance of the solutions found from
    the eigen-mode analysis.
This is done by asking whether the values given by the 
    solid lines in Fig.~\ref{fig_disp_diag}
    are present in time-dependent solutions.
We note that a similar study was conducted for step-function density profiles
    by \citeauthor{2005A&A...441..371T}(\citeyear{2005A&A...441..371T}, hereafter TOB05).
To start, we note that an equation governing the transverse velocity perturbation $\delta v_x(x, z, t)$
    can be readily derived from the linearized, time-dependent, cold MHD equations.
Formally expressing $\delta v_x(x, z, t)$ as 
    $v(x, t)\sin(kz)$, one finds that $v(x, t)$ is governed by (e.g., TOB05)
\begin{equation}
\label{eq_vx}
 \frac{\partial^2 v(x, t)}{\partial t^2}
    =v_{\rm A}^2(x)\left(\displaystyle\frac{\partial^2}{\partial x^2}
	-k^2\right)v (x, t)~.
\end{equation}
When supplemented with appropriate boundary and initial conditions,
    Eq.~(\ref{eq_vx}) can be readily solved such that 
    $v(x, t)$ at
    some arbitrarily chosen $x$ can be followed.
In practice, it is solved with a simple finite-difference code, second-order 
    accurate in both space and time, on a uniform grid spanning
    $[-x_{\rm outer}, x_{\rm outer}]$ with a spacing $\Delta x = 0.01R$
    and $x_{\rm outer} = 500R$.
For simplicity, we require that $v(x=\pm x_{\rm outer}, t) = 0$.    
A uniform time-step $\Delta t$ is
    chosen to be $0.8 \Delta x/v_{\rm Ae}$ in view of the Courant condition.
We have made sure that further refining the grid
    does not introduce any appreciable change.
In addition, the outer boundaries $x=\pm x_{\rm outer}$ are
    placed sufficiently far from the slab to ensure
    that the $v(x, t)$ signals are not contaminated by the perturbations reflected off
    these boundaries.

Consider kink modes first.
To this end, we adopt the following initial condition (IC)
\begin{eqnarray*}
\label{eq_num_ini_kink}
&& v(x, t)|_{t=0} = 
      \exp\left[-\left(\frac{x}{R/2}\right)^2\right]
     -\exp\left[-\left(\frac{x-R}{R/2}\right)^2\right]
     -\exp\left[-\left(\frac{x+R}{R/2}\right)^2\right], \\ 
&& \left.\displaystyle\frac{\partial v(x, t)}{\partial t}\right|_{t=0} = 0,
\end{eqnarray*}
    which represents an initial perturbation of even parity.
The solid line in Fig.~\ref{fig_timevol_kink} represents the temporal evolution of $v(0, t)$ 
    for a slab with an inverse-parabolic density profile.
Here we choose $\rho_{\rm i}/\rho_{\rm e}$ to be $10$ and 
    $l/R$ to be $1$.
In addition, a value of $0.1\pi$ is adopted for $kR$, 
    for which Figs.~\ref{fig_disp_diag}a and \ref{fig_disp_diag}b indicate that
    $\omega_{\rm R} = 0.616~v_{\rm Ai}/R$ for branch I,
    while $[\omega_{\rm R}, \omega_{\rm I}] = [3.52, -0.585]~v_{\rm Ai}/R$ for branch II.
The dashed line in red provides a numerical fit to the time-dependent solution with a function
    of the form $A\cos(\omega_{\rm R}+\phi)$ where the value for branch I is used for $\omega_{\rm R}$.
Evidently, the solid line for $t \gtrsim 10~R/v_{\rm Ai}$
    agrees remarkably well with the dashed line, meaning that the signal evolves into
    the trapped mode found in the eigen-mode analysis.
Furthermore, a Fourier analysis of the signal for $t \lesssim 10~R/v_{\rm Ai}$ 
    reveals a periodicity of $1.7 \sim 1.8$~$R/v_{\rm Ai}$,
    or equivalently, an angular frequency of $3.5 \sim 3.7~v_{\rm Ai}/R$.
This is in close accordance with the expectation from the first leaky mode labeled II,
    even though the signal decays too rapidly for one to 
     determine accurately the period and damping rate.
We note that this is why TOB05 named this stage ``the impulsive leaky phase'',
    since wave leakage plays an essential role in this evolutionary stage. 
When compared with Fig.~10 in TOB05, our Fig.~\ref{fig_timevol_kink} indicates
    that despite the quantitative differences, 
    the temporal evolution of kink perturbations for diffuse slabs is qualitatively 
    similar to what happens for slabs with a step-function form.

We now examine sausage modes by adopting the IC
\begin{eqnarray*}
\label{eq_num_ini_saus}
&& v(x, t)|_{t=0} = \frac{2x/R}{1+(x/R)^2}, \\ 
&& \left.\displaystyle\frac{\partial v(x, t)}{\partial t}\right|_{t=0} = 0,
\end{eqnarray*}
    which represents an initial perturbation of odd parity
    and localized around $x= \pm R$.
The temporal evolution of $v(R, t)$ is presented by the solid lines 
    in Fig.~\ref{fig_timevol_saus}, 
    where two different values of $kR$ are examined
    for a slab also characterized by an inverse-parabolic profile
    with $\rho_{\rm i}/\rho_{\rm e} = 10$
    and $l/R=1$.
The result for $kR = 0.01\pi$ is shown in Fig.~\ref{fig_timevol_saus}a,
    pertinent to the leaky mode with 
    $[\omega_{\rm R}, \omega_{\rm I}] = [1.83, -0.447]~v_{\rm Ai}/R$
    as expected from Figs.~\ref{fig_disp_diag}c and \ref{fig_disp_diag}d.
The red dashed line represents a fit to the time-dependent signal with 
    $A\cos(\omega_{\rm R} t +\phi)\exp(\omega_{\rm I} t)$,
    and is found to agree well with the solid line.
On the other hand, Fig.~\ref{fig_timevol_saus}b examines the case where $kR=0.25\pi$, 
    pertinent to the trapped mode with $\omega_{\rm R}=2.23~v_{\rm Ai}/R$
    as expected from Fig.~\ref{fig_disp_diag}c.
The red dashed line represents a fit to the time-dependent result with 
    $A\cos(\omega_{\rm R}+\phi)$.
Evidently, the red dashed line is hard to tell apart
    from the time-dependent result.
    
To summarize this section, we remark that the perturbations bearing signatures expected from
    the eigen-mode analysis can be readily generated.
While the results are shown only at some specific $kR$ for an inverse-parabolic profile with some specific combination of 
    $[l/R, \rho_{\rm i}/\rho_{\rm e}]$, 
    the same conclusion has been reached when we experiment     
    with all the considered profiles for a substantial
    range of $l/R$,  $\rho_{\rm i}/\rho_{\rm e}$, and $kR$.

\section{FAST WAVES IN NONUNIFORM SLABS WITHOUT A UNIFORM CORE}
\label{sec_App_DR}
In this section we will examine slabs with equilibrium density profiles of the form
\begin{eqnarray}
\label{eq_App_rhoprofile}
 {\rho}(x)=\left\{
   \begin{array}{ll}
   \rho_{\rm i}+(\rho_{\rm e}-\rho_{\rm i})f(\bar{x}),    & 0\le x < R, \\
   \rho_{\rm e},    & x \ge R ,
   \end{array}
   \right.
\end{eqnarray}
    where $\bar{x} = x/R$, and $f(\bar{x})$ is an arbitrary function that smoothly
    connects $0$ at $\bar{x}=0$
    to $1$ at $\bar{x}=1$.
Redefining $\zeta$ as $x-R/2$, the Fourier amplitude of the transverse Lagrangian displacement $\tilde{\xi}_x$
    can be expressed as 
\begin{equation}
\label{eq_App_xi}
   \tilde{\xi}_x(x)=\left\{
   \begin{array}{ll}
   A_1\tilde{\xi}_1(\zeta)+A_2\tilde{\xi}_2(\zeta),& 0\le x < R,\\[0.1cm]
   A_{\rm e}\exp(i\mu_{\rm e} x),    & x > R,
   \end{array}
   \right.
\end{equation}
    where $A_1$, $A_2$ and $A_{\rm e}$ are arbitrary constants,
    while $\tilde{\xi}_1$ and $\tilde{\xi}_2$ represent
    two linearly independent solutions to Eq.~(\ref{eq_xi}) for $x< R$.
Unsurprisingly, $\tilde{\xi}_1$ and $\tilde{\xi}_2$ are still describable by Eqs.~(\ref{eq_xi1xi2_expansion})
    to (\ref{eq_a_n}).
The Fourier amplitude of the Eulerian perturbation of the total pressure reads
\begin{equation}
\label{eq_App_pt}
   \tilde{p}_{\rm tot}(x)
  =-\displaystyle\frac{B^2}{4\pi}\times\left\{
   \begin{array}{ll}
   A_1 \tilde{\xi}_1^\prime(\zeta)+A_2\tilde{\xi}_2^\prime(\zeta),& 0 \le x < R,\\[0.1cm]
   iA_{\rm e}\mu_{\rm e}\exp(i\mu_{\rm e} x) ,    		  & x > R.
   \end{array}
   \right.
\end{equation}
     
The derivation of the dispersion relation (DR) follows closely the one given in Sect.~\ref{sec_sub_DR},
    the only difference being that in place of the interface between a uniform {\bf core} and the 
    transition layer, the slab axis $x=0$ needs to be examined.
To be specific, $\tilde{\xi}_x$ is required to be of even (odd) parity for kink (sausage) modes,    
\begin{equation}
\label{eq_App_x=0}
 \begin{array}{ll}
      A_1\tilde{\xi}_1(\zeta_{\rm i}) +A_2\tilde{\xi}_2(\zeta_{\rm i}) =0~, & {\rm sausage,}\\
      A_1\tilde{\xi}'_1(\zeta_{\rm i})+A_2\tilde{\xi}'_2(\zeta_{\rm i})=0~, & {\rm kink ,}
 \end{array}
\end{equation}
    where $\zeta_{\rm i} = -R/2$.
On the other hand, the continuity of both $\tilde{\xi}$ and $\tilde{p}_{\rm tot}$
    at $x=R$ means that
\begin{eqnarray*}\label{eq_ori2a}
  A_1\tilde{\xi}_1(\zeta_{\rm e}) +A_2\tilde{\xi}_2(\zeta_{\rm e}) &=&  A_{\rm e}\exp(i\mu_{\rm e} R) , \\ [0.1cm]
  A_1\tilde{\xi}'_1(\zeta_{\rm e})+A_2\tilde{\xi}'_2(\zeta_{\rm e})&=& iA_{\rm e}\mu_{\rm e}\exp(i\mu_{\rm e} R) ,
\end{eqnarray*}
    where $\zeta_{\rm e} = R/2$.
Eliminating $A_{\rm e}$, one finds that 
\begin{eqnarray}
\label{eq_App_x=R}
 \displaystyle\frac{A_1\tilde{\xi}'_1(\zeta_{\rm e})
                   +A_2\tilde{\xi}'_2(\zeta_{\rm e})}
                   {A_1\tilde{\xi}_1(\zeta_{\rm e})
                   +A_2\tilde{\xi}_2(\zeta_{\rm e})}&=& i\mu_{\rm e}~.
 \end{eqnarray}
The algebraic equations governing $A_1$ and $A_2$ then  
    follow from Eqs.~(\ref{eq_App_x=0}) and (\ref{eq_App_x=R}), 
\begin{equation}
 \label{eq_App_AA}
  \begin{array}{rcl}
    && \Lambda_1A_1+\Lambda_2A_2= 0 \\  && \Lambda_3 A_1+\Lambda_4 A_2= 0~,
  \end{array}
\end{equation}
    with
  \begin{equation}
  \label{eq_App_YY}
  \begin{array}{ll}
    \Lambda_1=\left\{\begin{array}{ll}
    \tilde{\xi}_1(\zeta_{\rm i})~,& {\rm sausage}\\
    \tilde{\xi}'_1(\zeta_{\rm i})~,& {\rm kink}
    \end{array}\right.
    & \Lambda_2=\left\{\begin{array}{ll}
    \tilde{\xi}_2(\zeta_{\rm i})~,& {\rm sausage}\\
    \tilde{\xi}'_2(\zeta_{\rm i})~,& {\rm kink}
    \end{array}\right. \\ [0.4cm]
     \Lambda_3=i\mu_{\rm e}\tilde{\xi}_1(\zeta_{\rm e})-\tilde{\xi}'_1(\zeta_{\rm e})~,
   & \Lambda_4=i\mu_{\rm e}\tilde{\xi}_2(\zeta_{\rm e})-\tilde{\xi}'_2(\zeta_{\rm e}) .
  \end{array}
\end{equation}
For $[A_1, A_2]$ not to be identically zero, one needs to require that
\begin{equation}
 \label{eq_App_DR}
   \Lambda_1 \Lambda_4-\Lambda_2 \Lambda_3=0~,
\end{equation}
   which is the DR governing fast waves supported by magnetic slabs with equilibrium density
   profiles described by Eq.~(\ref{eq_App_rhoprofile}).

What happens for a step-function form of the density profile?
This may take place, for instance, when $f(\bar{x})= \bar{x}^\mu$ with $\mu\rightarrow \infty$. 
In this case one finds that $\rho_0 = \rho_{\rm i}$ and $\rho_n = 0$ for $n\ge 1$.
Equation~(\ref{eq_a_n}) then leads to that $a_n =0$ for odd $n$, 
   whereas for even $n=2m$ ($m = 1, 2, \cdots$),
\begin{eqnarray*}
 a_{2m} = \left(-1\right)^m\frac{\mu_{\rm i}^{2m}}{(2m)!}a_0~,
\end{eqnarray*}
   meaning that $\tilde{\xi}_1$ can be expressed as
\begin{eqnarray}
   \tilde{\xi}_1 
 = a_0 \sum_{m=0}^\infty \left(-1\right)^m\frac{\mu_{\rm i}^{2m}}{(2m)!} \zeta^{2m}
 = a_0 \cos(\mu_{\rm i}\zeta) .
\end{eqnarray}
Note that $a_0 \ne 0$ but $a_1=0$.
Likewise, by noting that $b_0 = 0$ but $b_1 \ne 0$, one finds that 
    $b_n=0$ for even $n$, whereas for odd $n=2m+1$ ($m=1, 2, \cdots$),
\begin{eqnarray*}
 b_{2m+1} = \left(-1\right)^m\frac{\mu_{\rm i}^{2m}}{(2m+1)!} b_1~.
\end{eqnarray*}
As a result,
\begin{eqnarray}
   \tilde{\xi}_2 
 = b_1 \sum_{m=0}^\infty \left(-1\right)^m\frac{\mu_{\rm i}^{2m}}{(2m+1)!} \zeta^{2m+1}
 = \frac{b_1}{\mu_{\rm i}} \sin(\mu_{\rm i}\zeta) .
\end{eqnarray}
Evaluating the coefficients $\Lambda_n$ ($n=1, \cdots, 4$) in Eq.~(\ref{eq_App_DR})
   with the explicit expressions
   for $\tilde{\xi}_1$ and $\tilde{\xi}_2$, one finds that for sausage waves,
\begin{eqnarray*}
  0 &=& i\mu_{\rm e} [2\sin(\mu_{\rm i}R/2)\cos(\mu_{\rm i}R/2)]
         -\mu_{\rm i}[\cos^2(\mu_{\rm i}R/2)-\sin^2(\mu_{\rm i}R/2)] \\
    &=& i\mu_{\rm e}\sin(\mu_{\rm i}R)-\mu_{\rm i}\cos(\mu_{\rm i}R),
\end{eqnarray*}
   thereby recovering the step-function case (cf. Eq.~\ref{eq_DR_tophat}).
The DR for kink waves can be simplified in a similar fashion.
Finally, let us note that although $\tilde{\xi}_1(\zeta)$ ($\tilde{\xi}_2(\zeta)$)
    is an even (odd) function, 
    it is not so when $x=\zeta+R/2$
    is seen as the independent variable.
However, some algebra using trigonometric identities 
    shows that the combination of the two ($A_1\tilde{\xi}_1+A_2\tilde{\xi}_2$)
    is proportional to $\sin(\mu_{\rm i} x)$ ($\cos(\mu_{\rm i} x)$)
    for sausage (kink) waves, thereby explicitly showing
    the parity of the eigen-functions implied by Eq.~(\ref{eq_App_x=0}).

\begin{figure}
\centering
\includegraphics[width=0.85\columnwidth]{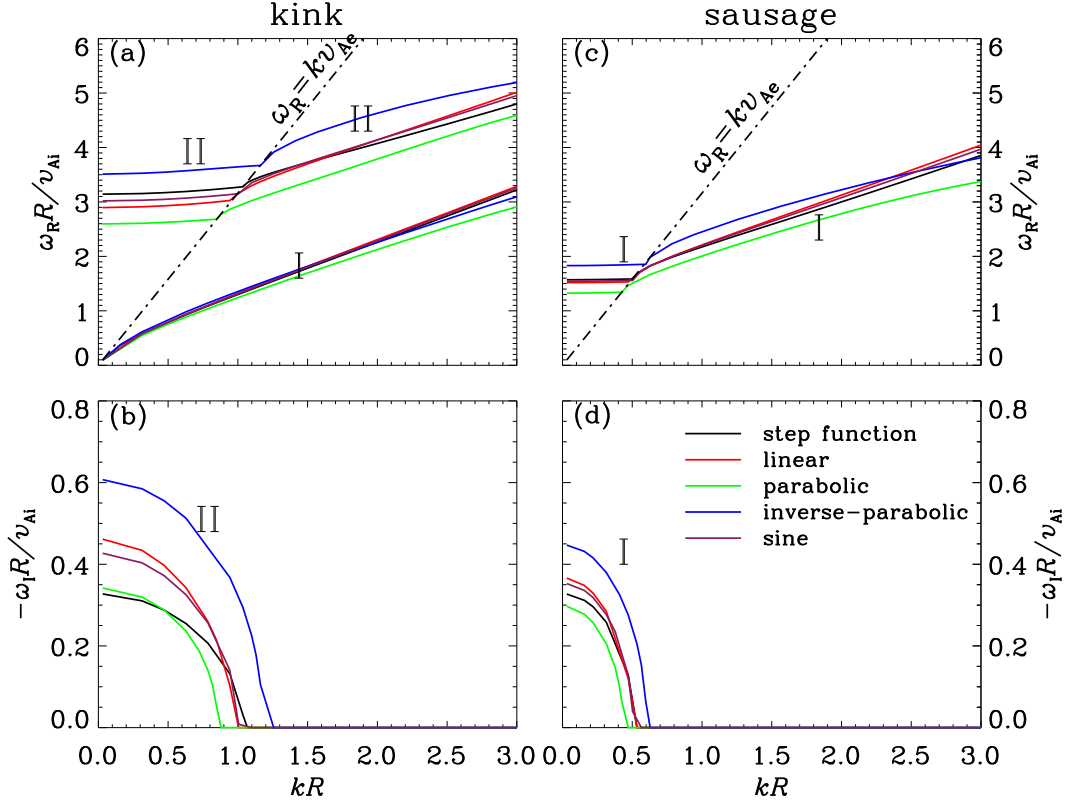}
 \caption{
  Dispersion diagrams for standing kink (left column) and sausage (right) modes in nonuniform slabs.
  The real ($\omega_{\rm R}$, the upper row) and imaginary ($\omega_{\rm I}$, lower row)
      parts of angular frequency
      are shown as functions of the real longitudinal wavenumber $k$.
  These solutions are found by solving the dispersion relation (Eq.~\ref{eq_DR})
      for four different density profiles as represented by
      the curves in different colors.
  The corresponding result for a step-function profile is given  
      by the black solid curves for comparison.
  In (a) and (c), the black dash-dotted lines represent $\omega_{\rm R} = k v_{\rm Ae}$ and separate
      the trapped (to its right) from leaky (left) regimes.
  Here the width of the transition layer $l=R$,
     and the density contrast $\rho_{\rm i}/\rho_{\rm e}=10$.
     }
 \label{fig_disp_diag}
\end{figure}

\begin{figure}
\centering
\includegraphics[width=0.85\columnwidth]{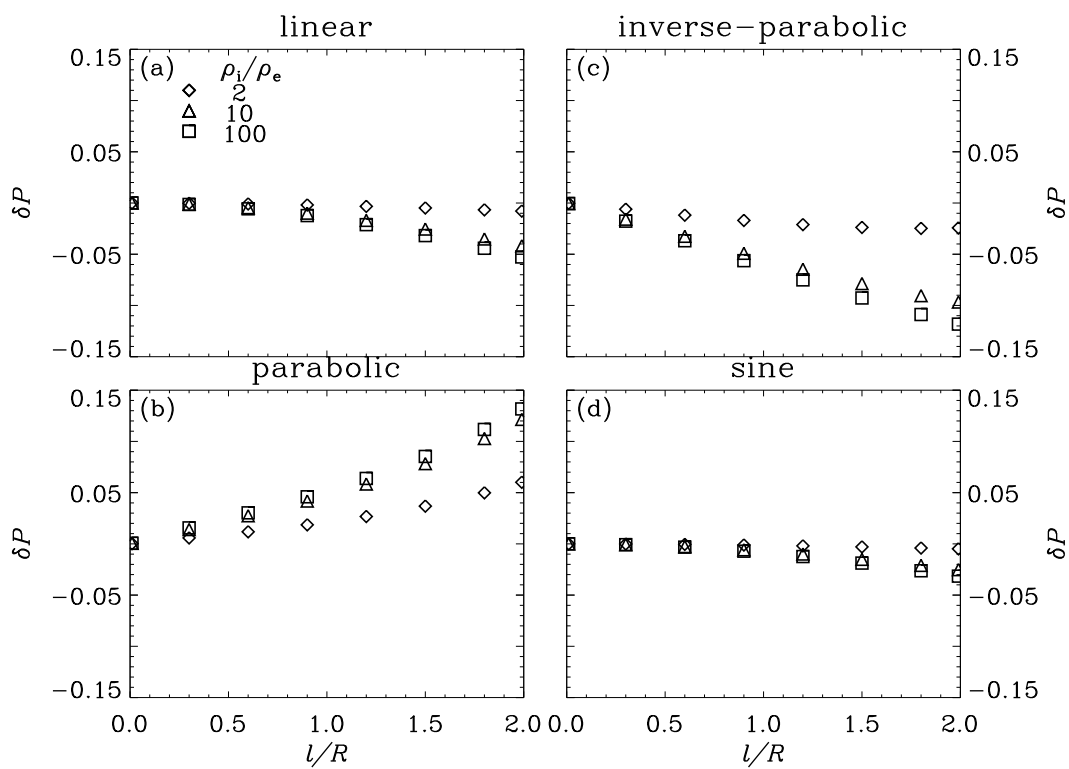}
 \caption{
 Influence of density profiles on periods $P$ of standing kink modes labeled I in Fig.~\ref{fig_disp_diag}.
 These modes are trapped at arbitrary longitudinal wavenumber $k$.
 Here $\delta P$ measures the most significant deviation of $P$
      from the step-function case 
      when $kR$ varies between $0$ and $0.2\pi$.
 A number of density contrasts are examined and represented by different symbols as labeled. 
  }
\label{fig_maxdevP_trappedkink}
\end{figure}

\begin{figure}
\centering
\includegraphics[width=0.65\columnwidth]{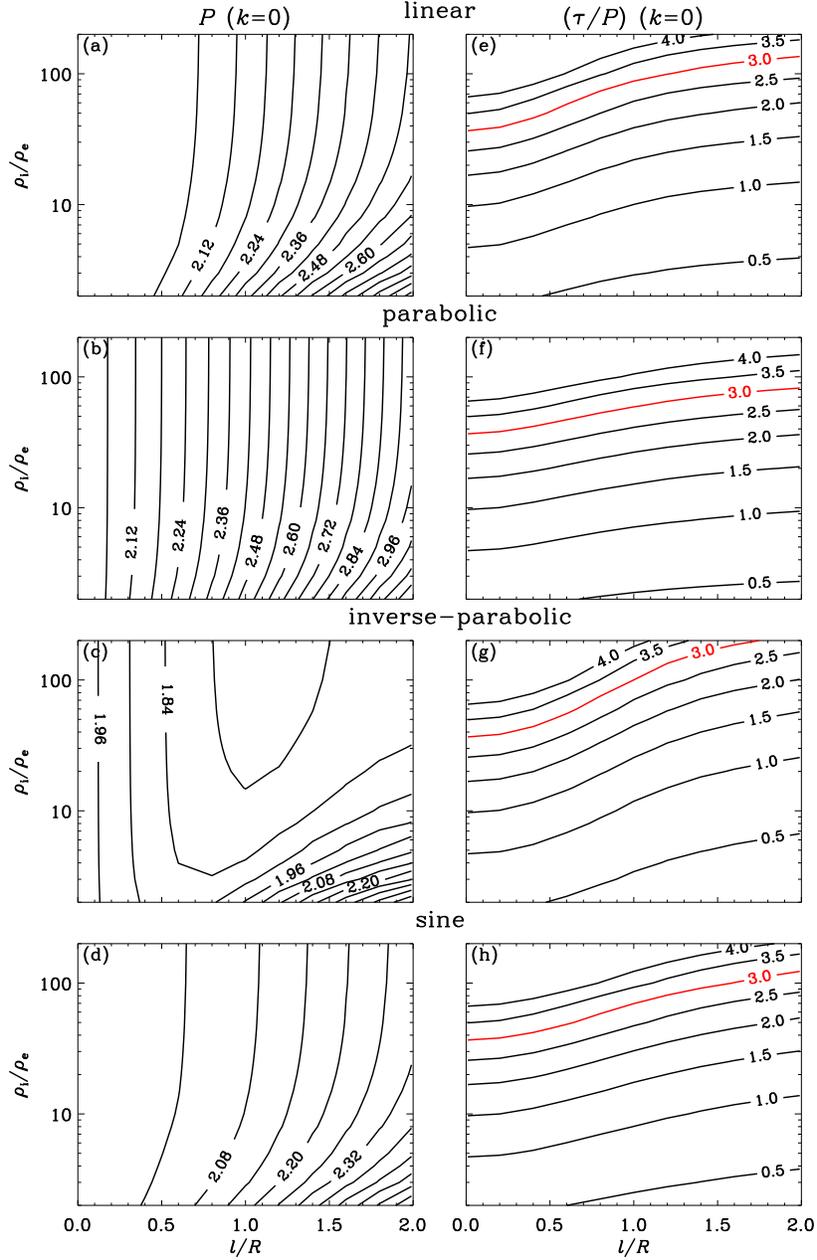}
 \caption{
 Influence of density profiles on periods $P$ (the left column) 
     and damping-time-to-period ratios $\tau/P$ (right)
     of standing kink modes labeled II in Fig.~\ref{fig_disp_diag}.
 These modes are leaky for small longitudinal wavenumbers $k$.     
 Contour plots in the $[l/R, \rho_{\rm i}/\rho_{\rm e}]$ space are shown 
     for $P$ and $\tau/P$ evaluated at $k=0$.
 Each row represents one of the four different density profiles as labeled.
 The red curves in the right column represent where $\tau/P=3$,
     the nominal value required for a signal to be observationally discernible.
  }
\label{fig_contour_kinkPtau}
\end{figure}

\begin{figure}
\centering
\includegraphics[width=0.65\columnwidth]{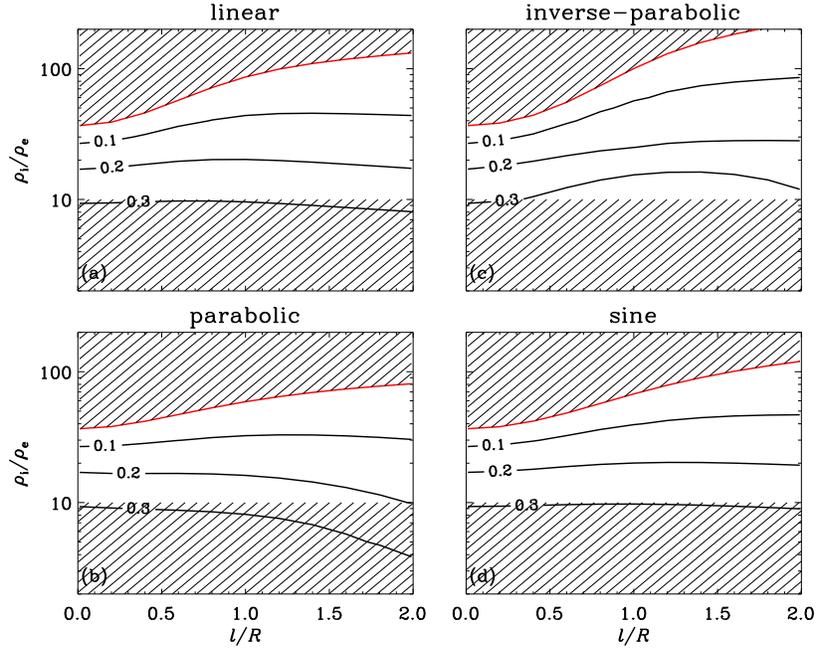}
 \caption{
 Observability of standing kink modes labeled II in Fig.~\ref{fig_disp_diag}.
 Contour plots in the $[l/R, \rho_{\rm i}/\rho_{\rm e}]$ space are shown
     for $(R/L)_{\rm obs}$ at which the damping-time-to-period ratio $\tau/P$
     attains three for a given pair of $[l/R, \rho_{\rm i}/\rho_{\rm e}]$.
 The red curve in each panel represents where $\tau/P$ at $k=0$ attains three.
 In the hatched area bounded from below by this curve, standing kink modes labeled II
     have sufficiently high quality to be observable, regardless of the geometrical size
     of slabs.
 The lower hatched area represents where $\rho_{\rm i}/\rho_{\rm e} \le 10$, representative of
     density contrasts of active region loops.
 See text for details.
     }
\label{fig_RoLobs_kink}
\end{figure}

\begin{figure}
\centering
\includegraphics[width=0.65\columnwidth]{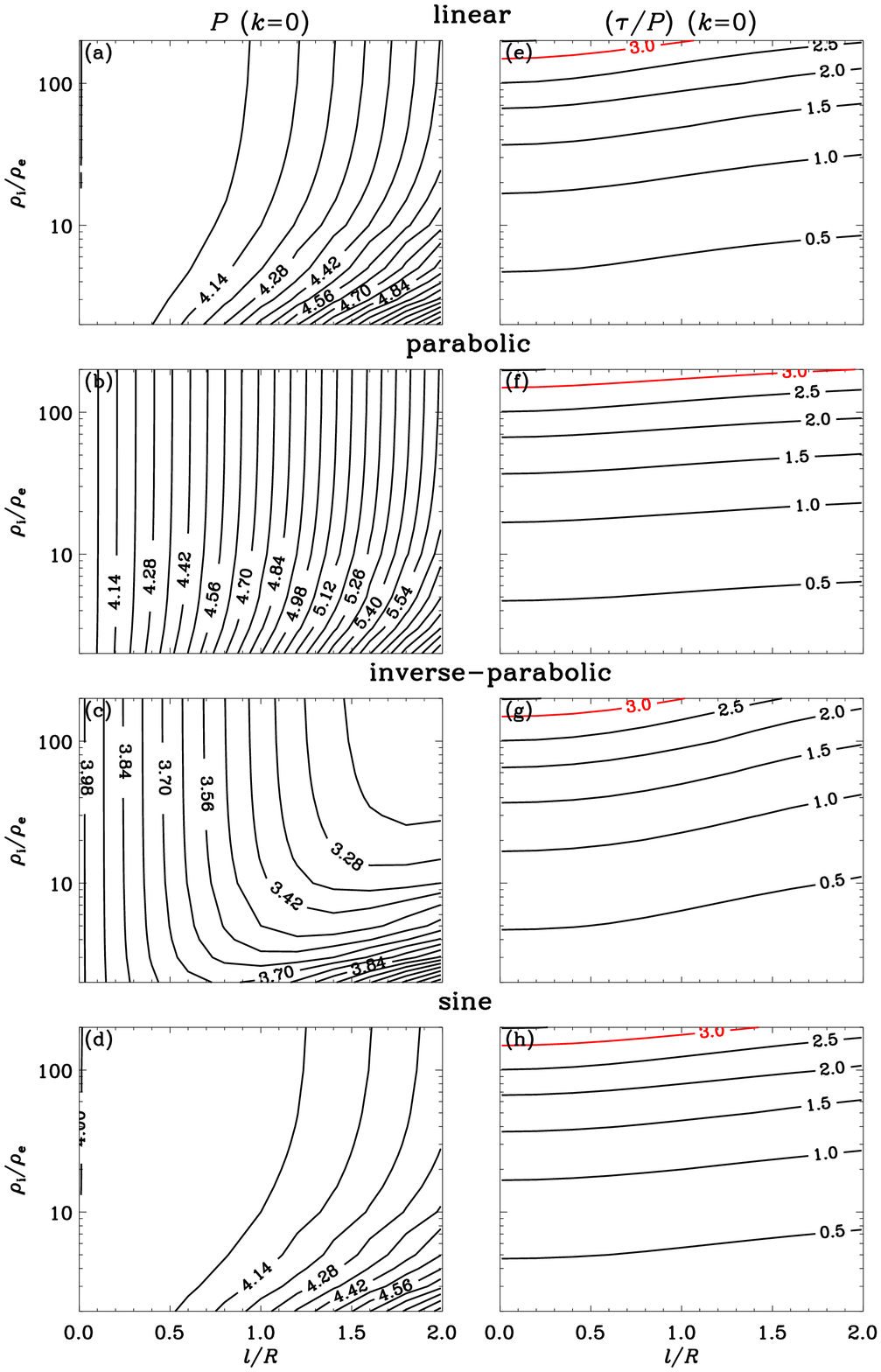}
 \caption{
 Similar to Fig.~\ref{fig_contour_kinkPtau} but for standing sausage modes.
 }
\label{fig_contour_sausPtau}
\end{figure}

\begin{figure}
\centering
\includegraphics[width=0.85\columnwidth]{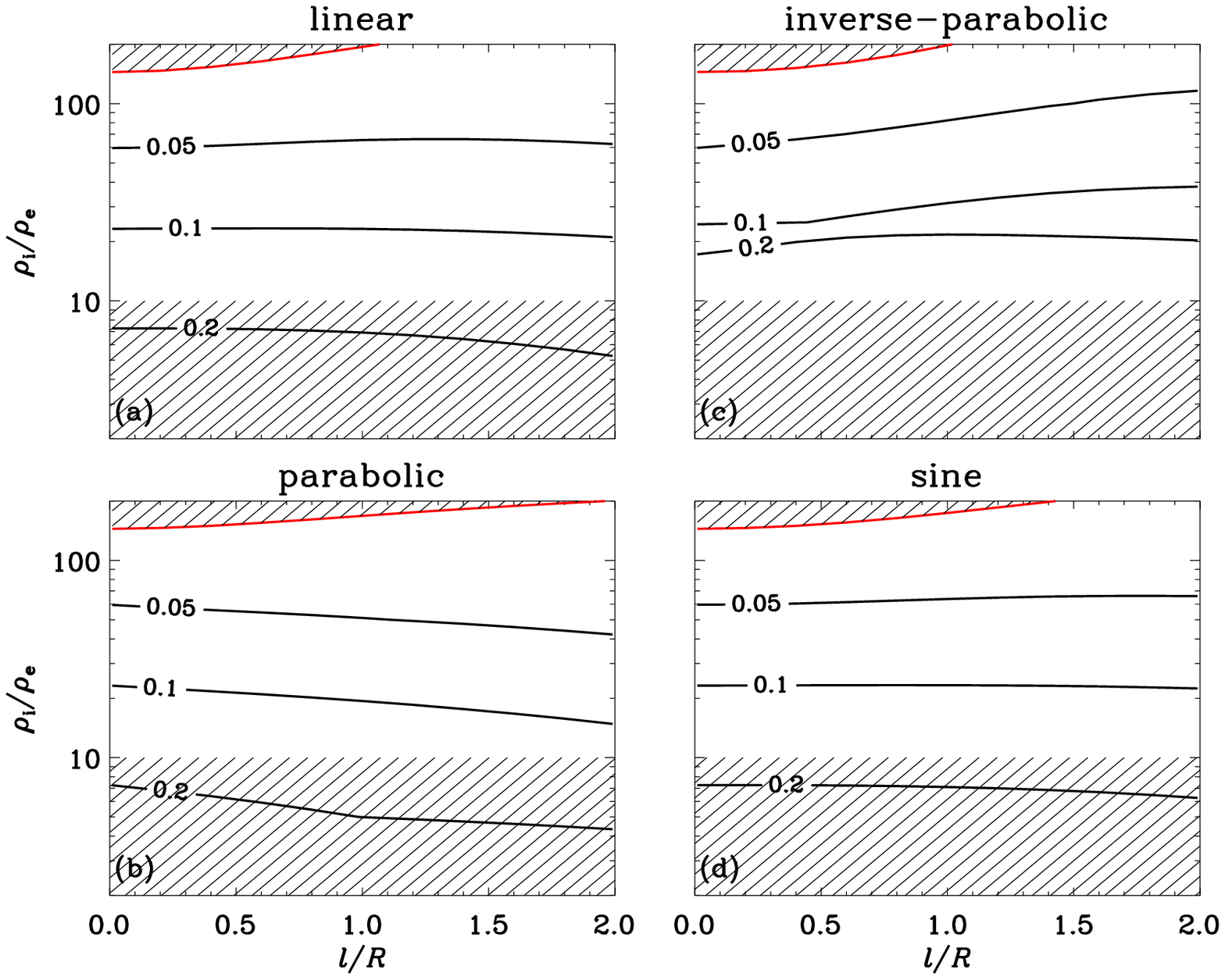}
 \caption{
Similar to Fig.~\ref{fig_RoLobs_kink} but for standing sausage modes.
  }
\label{fig_RoLobs_saus}
\end{figure}

\begin{figure}
\centering
\includegraphics[width=0.85\columnwidth]{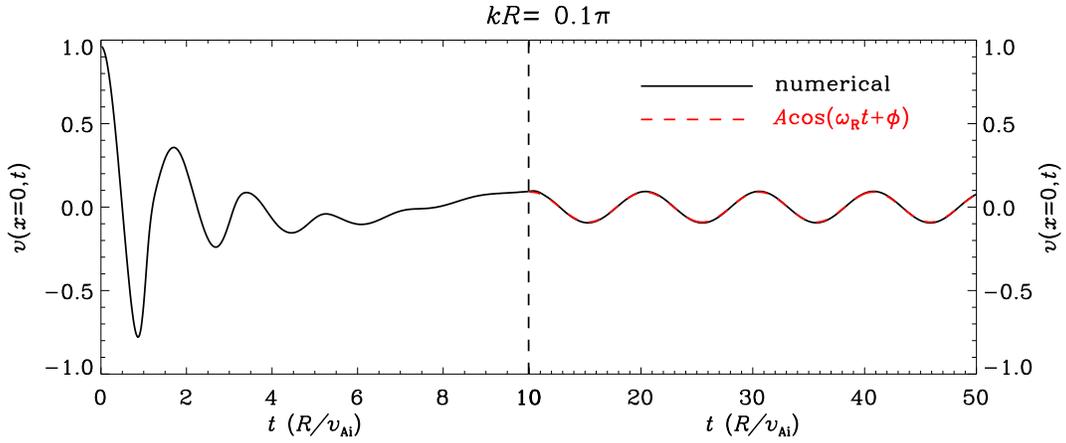}
 \caption{
Temporal evolution of transverse velocity perturbation
    $v(x=0, t)$ (the black curve) associated with kink modes
    supported by a slab with an inverse-parabolic density profile characterized by
    $\rho_{\rm i}/\rho_{\rm e} = 10$ and $l/R = 1$.
Here the longitudinal wavenumber $k=0.1\pi/R$.
The red dashed curve represents a fit to the time-dependent solution
    of the form $A\cos(\omega_{\rm R} t +\phi)$ where $\omega_{\rm R}$ is 
    found from the eigen-mode solution labeled I in Fig.~\ref{fig_disp_diag}a.
A different scale is used for the horizontal axis for $t<10 R/v_{\rm Ai}$.
  }
\label{fig_timevol_kink}
\end{figure}

\begin{figure}
\centering
\includegraphics[width=0.85\columnwidth]{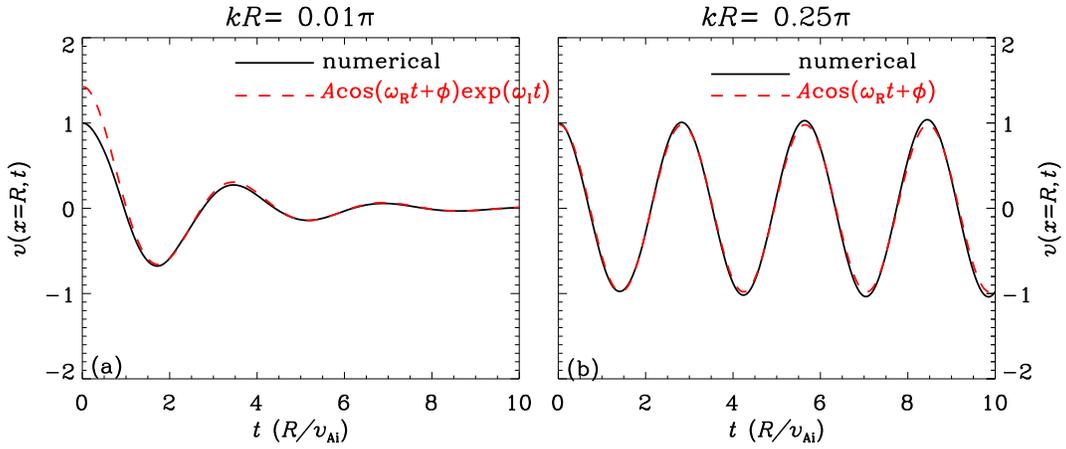}
 \caption{
Temporal evolution of transverse velocity perturbations $v(x=R, t)$ (the black curves) 
    associated with sausage modes
    supported by a slab with an inverse-parabolic density profile characterized by
    $\rho_{\rm i}/\rho_{\rm e} = 10$ and $l/R = 1$.
The results for two different values of the longitudinal wavenumber $k$, 
    one $0.01\pi/R$ and the other $0.25 \pi/R$,
    are given in (a) and (b), respectively.
The red dashed curves represent a fit in the form $A\cos(\omega_{\rm R} t +\phi)\exp(\omega_{\rm I}t)$
    where the values for $[\omega_{\rm R}, \omega_{\rm I}]$ are obtained from
    the eigen-mode analysis presented in Figs.~\ref{fig_disp_diag}c and \ref{fig_disp_diag}d.
Note that $\omega_{\rm I} = 0$ when $kR = 0.25\pi$.    
  }
\label{fig_timevol_saus}
\end{figure}

For simplicity, we have explored only one specific $f(\bar{x})$, namely $\bar{x}^\mu$ with
   $\mu$ positive.
Rather than further presenting dispersion diagrams showing
   the solutions to Eq.~(\ref{eq_App_DR}),
   let us remark that these solutions behave in a manner
   qualitatively similar to the ones presented in Fig.~\ref{fig_disp_diag}.
And their physical relevance can also be demonstrated by time-dependent computations.


\pagebreak

\pagebreak

\pagebreak

\pagebreak

\pagebreak

\pagebreak

\pagebreak

\clearpage

\clearpage

\end{document}